\documentclass[conference]{IEEEtran}
\IEEEoverridecommandlockouts
\usepackage{cite}
\usepackage{amsmath,amssymb,amsfonts}
\usepackage{algorithmic}
\usepackage{graphicx}
\usepackage{textcomp}
\usepackage{xcolor}
\def\BibTeX{{\rm B\kern-.05em{\sc i\kern-.025em b}\kern-.08em
    T\kern-.1667em\lower.7ex\hbox{E}\kern-.125emX}}
\begin{document}

\title{Keep Your Stakeholders Engaged: Interactive Vision Videos in Requirements Engineering
{\footnotesize }
}

\author{
	\IEEEauthorblockN{
		Lukas Nagel
	}
	\IEEEauthorblockA{ 
		Leibniz University Hannover\\
		Software Engineering Group\\
		Hannover, Germany\\ 
		Email: lukas.nagel@inf.uni-hannover.de
	}
	\and
	\IEEEauthorblockN{
		Oliver Karras
	}
	\IEEEauthorblockA{
		TIB -- Leibniz Information Centre for\\Science and Technology\\
		Hannover, Germany\\ 
		Email: oliver.karras@tib.eu
	}
}

\maketitle
\begin{abstract}
One of the most important issues in requirements engineering (RE) is the alignment of stakeholders' mental models. Making sure that all stakeholders share the same vision of a changing system is crucial to the success of any project. Misaligned mental models of stakeholders can lead to conflicting requirements. A promising approach to this problem is the use of video showing a system vision, so-called vision videos, which help stakeholders to disclose, discuss, and align their mental models of the future system. However, videos have the drawback of allowing viewers to adopt a passive role, as has been shown in research on e-learning. In this role, viewers tend to be inactive, unfocused and bored while watching a video. In this paper, we learn and adopt findings from scientific literature in the field of e-learning on how to mitigate this passive role while watching vision videos in requirements engineering. In this way, we developed concepts that incorporate interactive elements into vision videos to help viewers stay focused. These elements include questions that are asked during the video and ways for viewers to decide what happens next in the video. In a preliminary evaluation with twelve participants, we found statistically significant differences when comparing the interactive vision videos with their traditional form. Using an interactive vision videos, viewers are noticeably more engaged and gather more information on the shown system.



\end{abstract}

\begin{IEEEkeywords}
vision video, interactivity, annotation, player, requirements engineering, active video watching, e-learning
\end{IEEEkeywords}

\section{Introduction}



The negotiation of requirements from stakeholders is one of the key factors for the success of any project \cite{van2000requirements}. However, stakeholders can have different mental models of the developed product's functionality \cite{glinz2015shared}. Creating a shared understanding between the stakeholders is crucial to a project's success, as requirements expressed by stakeholders with unaligned mental models can lead to conflicts during development \cite{van2000requirements}. Vision videos have been found to be one way to support achieving such a shared understanding \cite{karras2021supporting, karras2018software,karras2017enriching}. Project members can directly elicit feedback from stakeholders by presenting the current vision of the aspired system \cite{schneider2019refining, busch2020vision, karras2020using}. In this way, they can quickly identify misunderstandings. 



One major drawback of using videos in this manner is that viewers can adopt a passive role when consuming them \cite{chi2014icap, broll2007using}. Research on educational psychology has shown that a passive consumption of information is less desirable than active processing \cite{takashima2004model, chi2014icap}. For the context of vision videos, stakeholders who passively consume the video are especially problematic. An unfocused stakeholder might miss significant misunderstandings, thereby undermining the core purpose of vision videos.
For this reason, we need methods to keep viewers engaged and active. while watching vision videos in requirements engineering. Ideally, these methods work in an unintrusive manner to introduce as little disruption as possible to the viewing experience. The research field of e-learning  \cite{zhang2006instructional, vural2013impact, kolaas2015application} already has an extensive body of knowledge on concepts designed to keep the viewer engaged and active.

In this paper, we learn and adopt findings from this body of knowledge in the research field of e-learning to vision videos in requirements engineering.
Methods like posing questions regarding the video's content\cite{vural2013impact, kolaas2015application} or allowing viewers to move freely along the timeline of the video\cite{zhang2006instructional} are explored in the new context of vision videos making them interactive. We present a preliminary evaluation of these concepts with twelve participants. In this evaluation, we compare a prototypical implementation of the described ideas with a traditional video player. Our results indicate clear advantages for users of our prototype like a better understanding of the video's contents and a noticeably higher engagement of the viewers.

The rest of this paper is structured as follows: We present related work in section \ref{sec:rw}. Section \ref{sec:concepts} contains the main concepts applied to vision videos, before section \ref{sec:eval} presents the performed evaluation. Our results are laid out in section \ref{sec:results} and discussed in section \ref{sec:discussion}. Section \ref{sec:conclusion} concludes the paper.

\section{Related Work}
\label{sec:rw}

Various works have discussed the use of interactivity in videos. A paper by Shipman et al. \cite{shipman2003generation} describes automatically generated video summaries that can be interacted with. The interactions allow viewers to select the amount of detail included in the summary and to navigate to the source video. Their concepts are included in Hyper-Hitchcock \cite{shipman2003hyper}, a player and editor for such video summaries. 
Meixner et al. \cite{meixner2010siva} discuss \textit{SIVA Suite}, a tool for the creation and playing of interactive non-linear videos. Within this tool, videos can be cut and enriched with a scene graph and fork nodes enabling a choice between alternative paths. They propose the usage of their system in e-learning contexts.

Other related research focused on the engagement of viewers. Galster et al. \cite{galster2018toward} use active video watching to educate software engineering students on soft-skills. They engage their students with the video's content by integrating interactive methods like commenting on videos, questions or discussions in peer groups. Results of an experiment making use of active video watching indicate that active discussions can enhance the learning experience. A work by Acharya et al.\cite{acharya2017using} also presents the use of case study videos to aid the education of software engineering students. They developed six hours of case study videos and received positive feedback on the effectiveness of case study videos as active learning tools.


The use of videos in requirements engineering has also been discussed in scientific literature \cite{Schneider.2017,karras2018software2}. Glinz \cite{glinz2000improving} defines scenarios as the description of requirements as sequences of interactions and discusses their potential for the requirements engineering process. Vision videos are based on presenting such scenarios within a video \cite{karras2021supporting}. According to Karras et al.\cite{karras2020representing}, vision videos are videos which represent a vision for the purpose of achieving shared understanding among all parties involved. Creighton et al.\cite{creighton2006software} present research on video-based requirements engineering. Their tool \textit{Xrave} shows scenarios in vision videos by arranging its contents in a scenario graph. The resulting multi-path scenes can be mapped to UML use cases. The paper is concluded with an outline of multiple challenges solved by the use of videos in requirements engineering, including differing perceptions of reality and gaps between stakeholder's terminology. Karras et al. \cite{karras2017video} developed an approach to create vision videos as a by-product of digital prototyping to enrich textual scenario descriptions. Their evaluation showed that this kind of videos support a faster understanding of the a textual scenario compared to static mockups. Karras et al. \cite{karras2017video} assumed that this difference results from the viewers' freedom to switch more easily between the video and the text than between the text and the static mockups.




In this paper we present concepts for an interactive video player, specifically in the context of vision videos used during the requirements negotiation phase. We base these concepts on research in the field of e-learning.

\section{Concepts}
\label{sec:concepts}

The overall goal of this research is to \textit{engage stakeholders with vision videos through active information processing}. To achieve this goal, we look to adopt findings research on e-learning that also relate to videos. We divide the shown video into sections consisting of individual scenes. Some interactions only appear in between such sections while others are present throughout the video. The following concepts look to keep stakeholders active while watching a vision video by offering or sometimes mandating interactions with the video player.

\subsection{Questions of Understanding}
\begin{figure}[!h]
	\centering
	\includegraphics[width=0.9\columnwidth]{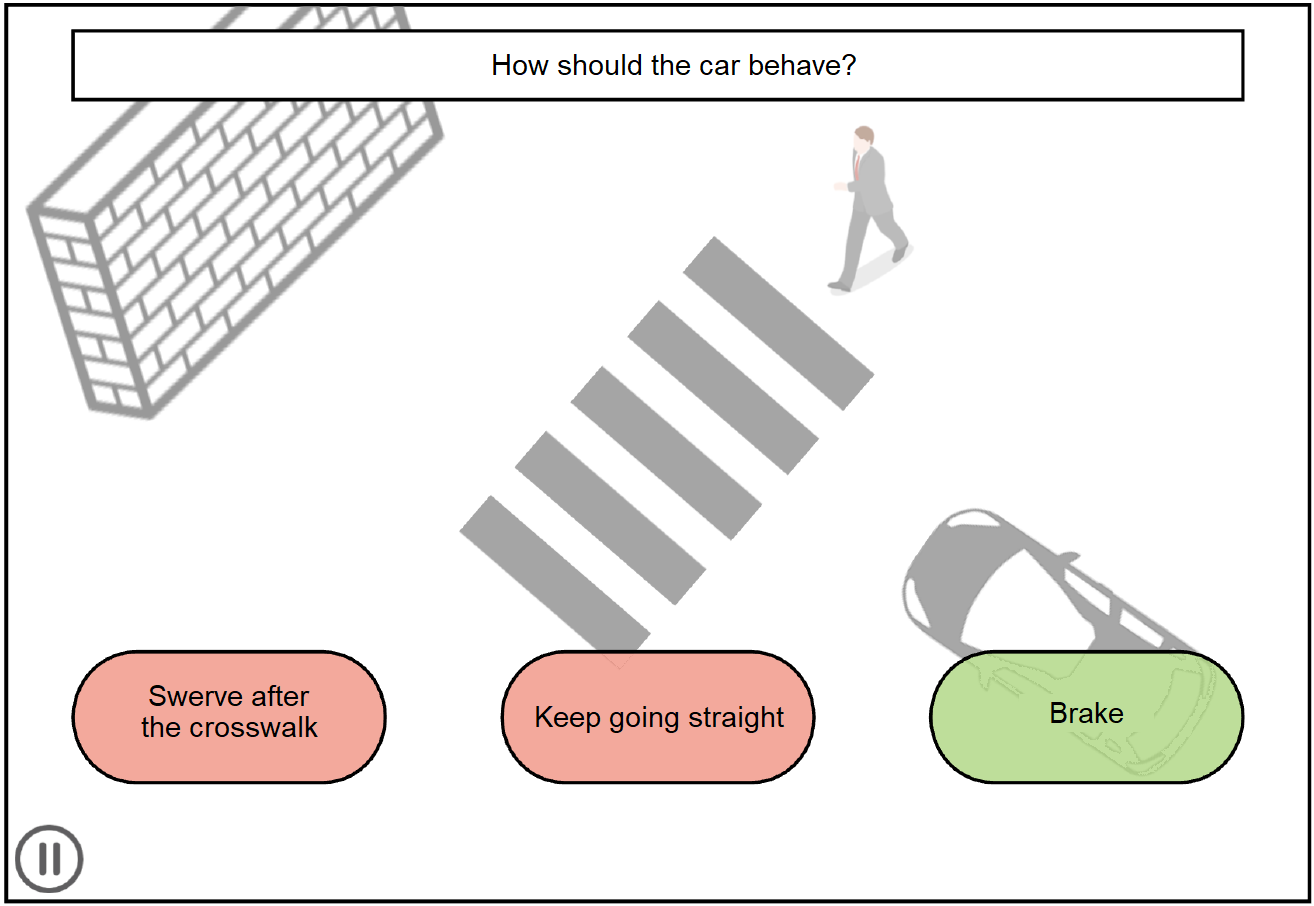}
	\caption{An example of a question being asked during a video.}
	\label{fig:questions}
\end{figure}

Questions regarding a viewers understanding of what is presented on screen have been used in the field of e-learning\cite{vural2013impact, kolaas2015application}. Multiple works\cite{vural2013impact, cummins2015investigating} have found the addition of questions to the video timeline to be preferable to asking them at the end. In contrast to the use of videos in e-learning contexts, however, we do not look to increase the learning effect. Rather, we seek to make ensure a consistent domain knowledge among stakeholders. Therefore, questions are asked immediately after the relevant video section has been shown. \figurename{ \ref{fig:questions}} presents an example for what the screen looks like when a question is asked. 

When a user answers the question, we mark the intended answer in green and the alternatives in red, before moving on with the content of the video. This design differs from questions in e-learning contexts, where wrongly answered questions are repeated. We pause the video when a question is asked. This means that there is no time pressure on the viewer.
Designing the questions in this way means that answers are more likely to be well thought out, instead of impulsive. Furthermore, answers to these questions do not impact the content shown in the video. 

\subsection{Alternative Paths}
\begin{figure}[h]
	\centering
	\includegraphics[width=\columnwidth]{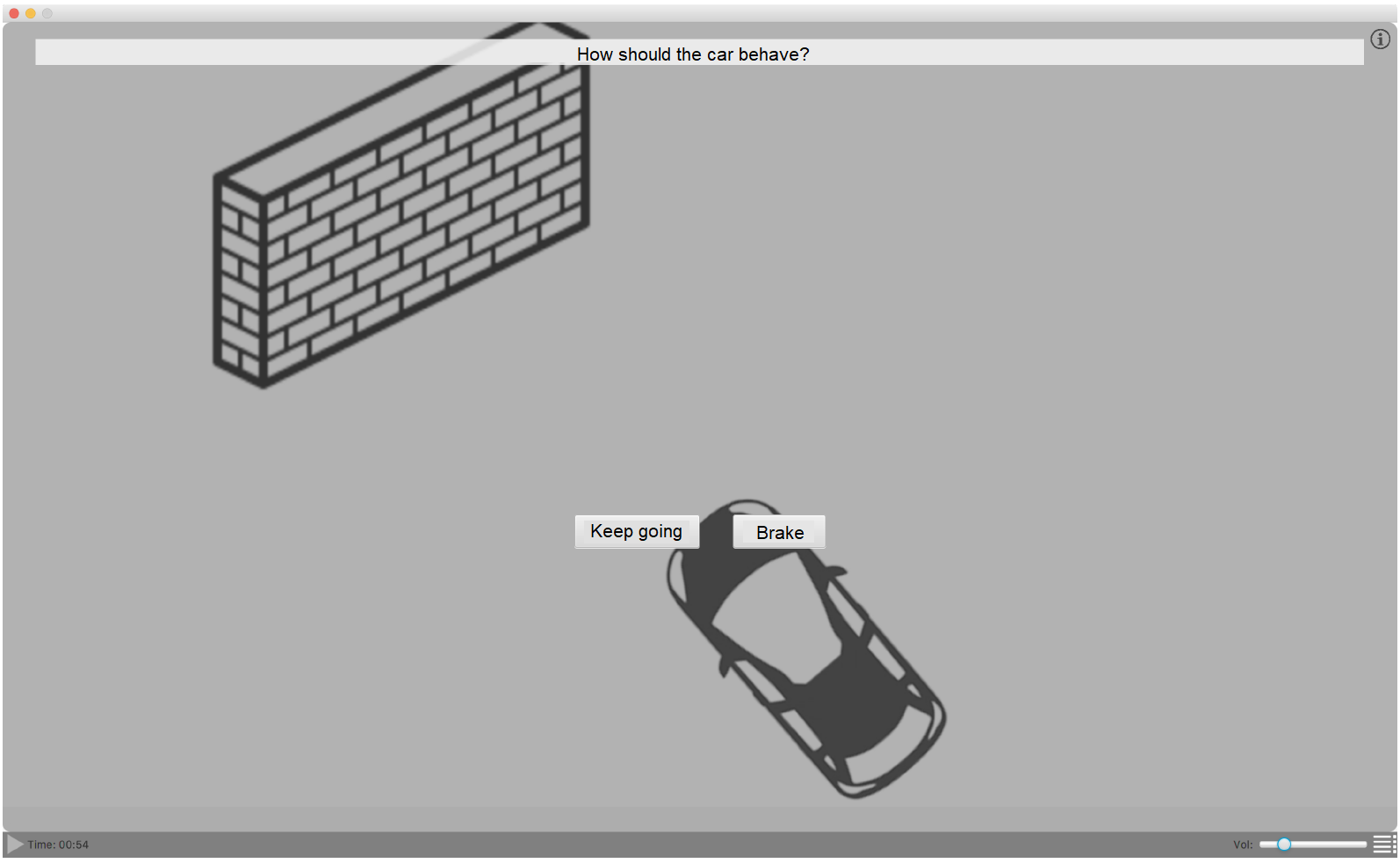}
	\caption{An example of a path choice. The user can decide how the car should behave in the next video section.}
	\label{fig:path}
\end{figure}

One of the core concepts of this paper is the offering of multiple paths. With vision videos often being used in early project stages, when the exact vision of the final product is not yet clear, presenting multiple alternative paths to a user can be beneficial. Through the choice of specific paths, various possibilities can be explored before making an informed decision on which path should be taken for the project. An example is presented in \figurename{ \ref{fig:path}}.

Alternative paths have already been used for vision videos\cite{creighton2006software, schneider2019refining}. In accordance to their work we did not include a timer forcing users to make a decision within a given time frame. Important decisions on the content of vision videos should not be made without thoughtful intent. Therefore, we design the choice of alternative paths with a pause in the video. This forces the viewer to make a decision and actively interact with the video. Additionally, they have to analyze the different choices and think about which path they prefer. Thereby, alternative paths present one way to engage stakeholders with the vision video. Questions of understanding as described are related to the content of individual paths. This means that the choice of a path impacts which question is asked. 

\subsection{Scene Overview}
With multiple scenes and different paths present in the video, a navigation menu of all available scenes is helpful. A creator of an interactive vision video can designate any scene or interaction as a navigation point. The viewer is shown a list of all designated navigation points including a timestamp, a title and its category, namely whether its a new scene, an alternative path or a question. While interacting with this overview, the video itself is paused, as the viewer is assumed to be busy reading its contents.
The inclusion of a scene overview incentivises viewers to revisit scenes of interest or decision points. The navigation menu also helps with a reflection of already viewed scenes, as references to all navigation points are included. Thereby, the overview helps viewers to adopt a more active role while watching the video.

\subsection{Annotations}
\begin{figure}[h]
	\centering
	\includegraphics[width=0.9\columnwidth]{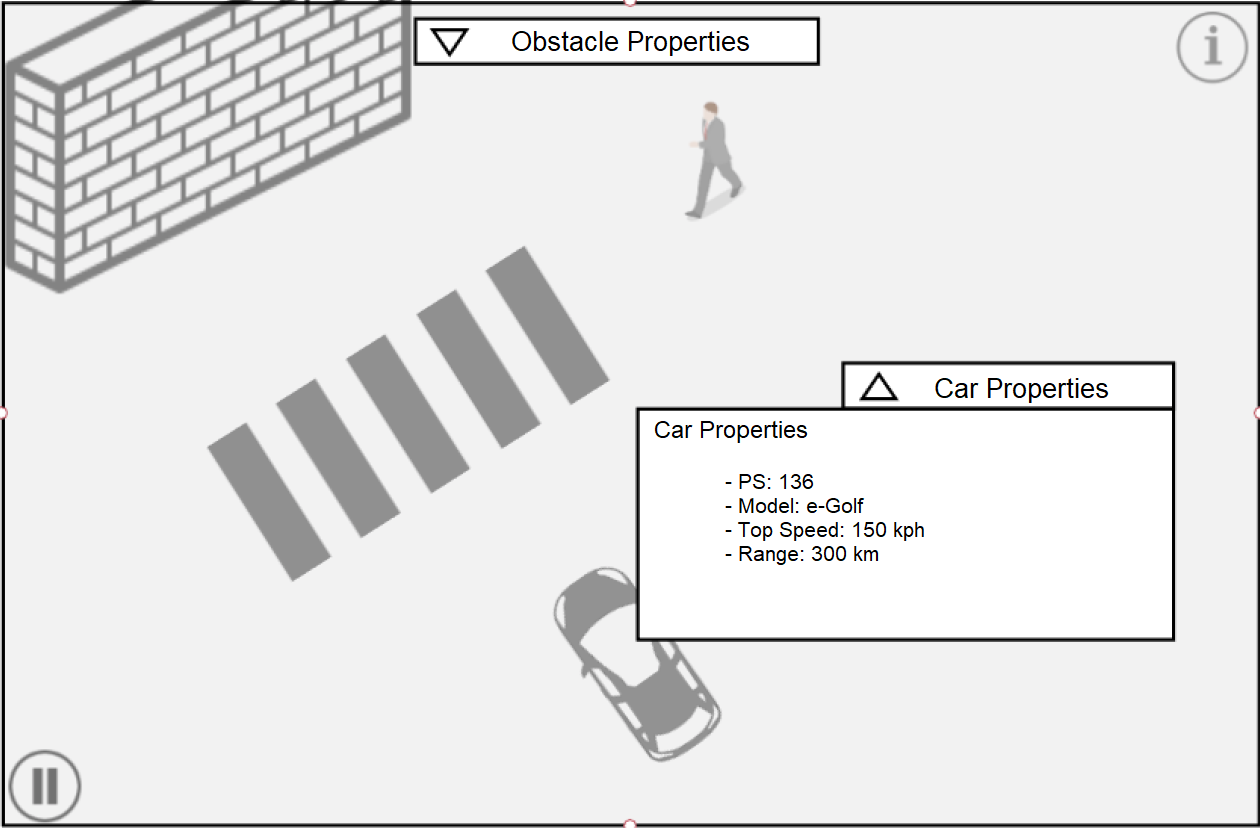}
	\caption{Annotations shown on the screen. The annotation titled \textit{Car Properties} is expanded.}
	\label{fig:annotations}
\end{figure}

Annotations enhance vision videos by providing additional information to the content of a video \cite{Karras.2016}. As shown in \figurename{ \ref{fig:annotations}}, we visualize them as small boxes with a title that can then be expanded by users. Expanded annotations show text, hyperlinks, images or even files related to the video's content. Including annotations provides interested viewers with details on demand. We differentiate between two different kinds of annotation. They can either be added to a vision video by its creator or added by a viewer. For example, viewers can annotate portions of the video they found to be unclear, while the creator can provide initial annotations to give more information in addition to whats shown in the video.

In order to keep such annotations from cluttering the screen, we opt to omit them from the visualization initially. Viewers can press a button in the top right corner to make annotations visible. The button is grayed out when no annotations are available to indicate that it is currently disabled. 
In terms of video behavior, we chose to keep the video playing when the small box annotations are shown, but to pause it once a viewer expands them. This design is chosen in order to keep the video flow from being disrupted too much. The video pauses when an annotation is expanded. Information shown in the video can not be processed at the same time as the expanded annotation.

\subsection{Export Function}
While the main purpose of the included interactions is to fulfill our research goal of \textit{engaging stakeholders with vision videos through active information processing}, they also lead to the creation of other useful data.


Exporting and analyzing data regarding the questions asked can provide crucial information regarding unclear aspects of the shown system. This allows requirements engineers to react to these aspects during the negotiation phase.

Recording and exporting the choices of paths made by individuals is useful, as information can be gathered on the consensus of stakeholders. For example, differences in the mental models of stakeholders or choices that are made by most viewers can be analyzed. Such recordings also offer insights on which scenes viewers watched and explored the most. 
Decisions that are still controversial among stakeholders can also be discovered. This information can indicate areas of interest for further discussions. The export uses structures similar to the ones presented in Meixner and Kosch\cite{meixner2012interactive}.

Finally, annotations added by viewers can also be analyzed. Misunderstandings or questions regarding the contents of the vision video can be extracted and discussed. Annotations added by viewers can also support a requirements engineer with more information from stakeholders. For example, an annotation could contain what exactly the stakeholder had originally imagined differently from what was shown. This provides more content for important discussions regarding the envisioned system.

\subsection{Concept Usage}
An interactive video player including the presented concepts should be used by stakeholders individually. Questions of understanding and path choices in particular are not designed to be used under any time pressure. This individual use facilitates the use of the presented concepts for groups of stakeholders that are distributed among different locations or time zones.

\section{Evaluation}
\label{sec:eval}
We conducted a study with university students in order to investigate the concepts outlined in section \ref{sec:concepts}.

\subsection{Research Goal and Research Questions}
The main goal of the experiment was to evaluate our concepts by comparing a prototypical implementation of the interactions with a traditional video player. We therefore ask the following research questions:

\noindent
\textbf{RQ1} \textit{How do users of the two players differ regarding the number of correctly answered questions?} 
During the study we posed multiple questions regarding the video's contents. A difference in the amount of correctly answered questions would indicate one player leading to a better understanding of the same video due to a more active processing of information.\newline 
\noindent
\textbf{RQ2} \textit{How do users of the two players differ regarding the amount of time spent with the videos?} 
Interactions adopted from research on e-learning incentivise the user to spend time on other aspects than purely watching the vision video. An interactive player could therefore lead to test subjects requiring more time to watch the same video when compared to their counterparts using a traditional player.\newline
\noindent
\textbf{RQ3} \textit{How do users of the two players differ regarding the amount of optional interactions executed?} 
A larger number of optional interactions indicates that participants do not just consume the video but are active while watching. It also alludes to a larger interest in the presented video.

\subsection{Hypotheses \& Variables}
We derived three hypotheses from the aforementioned research questions. The evaluation presented in this section is designed to test the following statements as well as their direct opposites:\newline

\begin{small}
	\noindent
	\textbf{H1$_0$} There is no difference in the number of correctly answered questions. \newline
	\textbf{H2$_0$} There is no difference in the amount of time subjects spent with the videos. \newline
	\textbf{H3$_0$} There is no difference in the number of optional interactions executed by participants. \newline
\end{small}

Our experiment followed a between-subjects design due to the difficulty of creating two similarly complex videos. The independent variable of the experiment is the type of video player. We asked one half of participants to watch a vision video with a prototypical implementation of our concepts, while the remaining test subjects used a traditional video player. The latter half thereby acted as a control group.

Multiple dependent variables were examined. We measured the amount of correctly answered questions, the time that each participant spent with the video, as well as the number of optional interactions executed. Additionally, we gathered information on the amount of different paths that were seen by each test subject over the course of the study.

\subsection{Material}
The study was conducted using a traditional video player and a prototypical implementation of the presented concepts. All participants used the same laptop, keyboard and mouse. The vision video shown was the one used in Schneider et al.\cite{schneider2019refining} It deals with the ordering and delivery processes of products in rural areas and contains two sets of three variants. 

The first set presents three ways of ordering product. The first variant makes use of an app that requires users to take a picture of the product. Another method works by pressing of a specific button placed in the near vicinity of the product while the third consists of an automatic order placement triggered by a measurement of the product's fill level. 

The second set of variants concerns the manner in which the ordered product is delivered. One variant works by the customer picking up the package from a neighbor who had brought it to the rural area from a city. In the second method showed a delivery made by a drone while the third had a postman drop the package into the trunk of a parked car belonging to the customer.

\paragraph{Participants Selection}
A total of twelve participants aged between 19 and 31 years took part in the experiment. Half of the participants are undergraduate students, four are graduate students and two participants have already finished their studies. All but one participant are involved in computer science related lines of study, while the single other participant studies physics and mathematics. Participants do not need any specific knowledge in the field of computer science, since the target audience of vision videos are stakeholders of the envisioned systems. Höst et al.\cite{host2000using} found only minor differences between students and software professionals regarding their performance of relatively small assessment tasks. As our experiment is similar to an evaluation by Karras et al.\cite{karras2020representing}, students are suitable test subjects. 

\subsection{Experiment Procedure}
Participants took part in the study individually. After an initial greeting, each participant was assigned to one of the two video players. Then, a quick introduction to the specific player's functionality and interaction methods was given. 

Test subjects were explicitly told that they could move freely within the video's timeline. This meant that they could rewind to previously shown scenes or explore other paths.

Participants who worked with the interactive video player were shown various overlays with possible interactions while watching the video. In accordance with the concepts presented in section \ref{sec:concepts}, interactions of choosing a path or answering a question were designed as mandatory, while informative annotations or the use of an overview menu were optional. Participants were asked to voice comments at specific points in the video in accordance to the work of Schneider et al.\cite{schneider2019refining}.

The control group working with the traditional video player was given a printed list of all annotations. Questions regarding its contents were only asked at the end. However, participants were allowed to search the video for answers. The choice of different paths that could be taken was performed by researchers asking participants to choose a path before moving to the respective point on the video timeline. It was also mentioned that participants were allowed to rewind and select a different path should they choose to do so. Comments was handled the same way as with participants using the interactive player. We counted the reading of an annotation on the list as an optional interaction to make the data comparable to data collected with the interactive video player.

\paragraph{Data Collection}
We measured the time that each participant spent watching the video or interacting with it. Additionally, we collected the answers to questions and the number of paths that each participant saw. We also recorded the number of optional interactions that each participant executed. After participants had finished watching the video, demographic data and personal feedback was collected. 

\subsection{Data Analysis Procedures}
\label{sec:dataAnalysis}
All data sets were analyzed for normal distribution using the Shapiro-Wilk Test\footnote{We opted for the Shapiro-Wilk Test due to the small sample size.}. Then, we applied the t-test in cases where a normal distribution is present. Otherwise we opted for the Mann-Whitney U test to examine the statistical significance of differences between the two players. 

\section{Results}
\label{sec:results}

\subsection{Correctly Answered Questions}
Participants were asked a maximum of six questions over the course of the experiment, two of which were only shown when specific paths were selected. We compare the number of correctly answered questions of all participants in relation to the maximum of six questions, as participants were explicitly told about the option of rewinding the video to explore other paths. With the Shapiro-Wilk test indicating a normal distribution to be present within the data, we calculate the statistical significance using the t-test. The results of our experiment show that users working with the interactive video player answered significantly more questions correctly than those using a traditional player (t(5) = 3.50, p = 0.006). We therefore reject \textbf{H1$_0$} and accept \textbf{H1$_1$}.


\subsection{Time Spent}
Test subjects who worked with the interactive video player spent an average of 10:12 minutes watching the video, with a median of 10:07 minutes and a standard deviation of 1:26 minutes. In contrast, the other group spent an average of 8:30 minutes, with a median of 8:42 minutes and a standard deviation of 00:46 minutes. The Shapiro-Wilk Test indicated a normal distribution. We applied the t-test to the data set and found no statistically significant difference (t(11) = -0.09, p = 0.926). This result aligns with our initial hypothesis \textbf{H2$_0$}. 

\subsection{Paths Seen}
Over the course of the experiment, each participant had to watch at least two paths to finish a video. However, a total of six individual paths were available. Participants using the traditional video player watched an average of 3.2 paths, with a median of 3. As for the interactive player, an average of 5.2 paths was seen, with a median of 5.5. The standard deviation for both sets of data was 0.98. Once again, we tested the data for a normal distribution, which was confirmed by the Shapiro-Wilk test. Therefore, we applied the t-test which indicated a statistically significant difference (t(5) = 3.52, p = 0.006).

\subsection{Optional Interactions}
\begin{figure}[h]
	\centering
	\includegraphics[width=0.9\columnwidth]{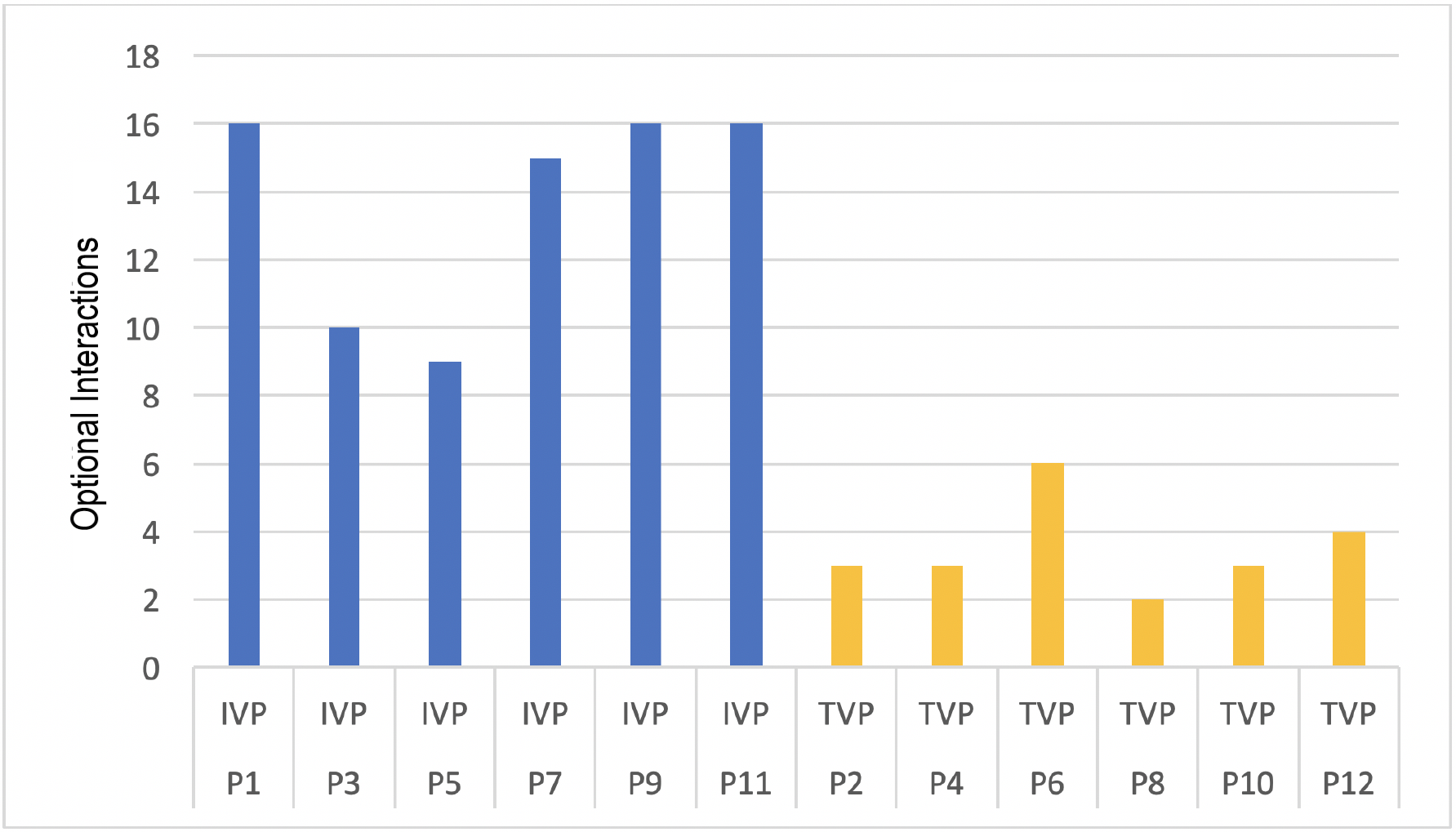}
	\caption{Amounts of optional interactions executed by each user. \textit{IVP} refers to the interactive video player, while \textit{TVP} refers to a traditional video player.}
	\label{fig:optionals}
\end{figure}

For the purposes of this paper we classify the opening of an annotation, the choice of an entry in the contents menu, as well as the viewing of an additional path as optional interactions. Users of the traditional video player opted for 3.4 of these interactions on average, with a median and standard deviation of 3 and 1.37 interactions respectively. As for the interactive player, 13.7 optional interactions were executed on average, with a median of 15.5 and a standard deviation of 3.27. Individual results of all participants can be found in \figurename{ \ref{fig:optionals}} The Shapiro-Wilk test found no normal distribution in the data. We used the Mann-Whitney U test and found a statistically significant difference (U = 0, p = 0.005). Due to this result we can not accept \textbf{H3$_0$} and accept \textbf{H3$_1$} instead.

\subsection{Comments}
Participants working with the prototypical implementation of the presented concepts wrote down 3 comments on average with a median of 3 and a standard deviation of 1.55. The control group provided 2.3 comments on average with a median of 1.5 and a standard deviation of 1.75. The Shapiro-Wilk test did not indicate a normal distribution. No statistically significant difference could be found with the Mann-Whitney U test (U = 13, p = 0.472).

\section{Discussion}
\label{sec:discussion}
We conclude this paper by discussing the results of our evaluation regarding our overall goal of \textit{including stakeholders in vision videos through active information processing}. We also discuss the threats to validity. For our research questions we found the following:

\noindent
\textbf{Answer to RQ1:} The acceptance of \textbf{H1$_1$} means that we found a statistically significant difference in the number of correctly answered questions. More specifically, our data indicates that users of the prototypical implementation of our concepts gave correct answers more often.


\noindent
\textbf{Answer to RQ2:} Based on the results of our experiment we could not reject \textbf{H2$_0$}. We found no statistically significant difference in the times test subjects spent watching the video.

\noindent
\textbf{Answer to RQ3:} Data gathered over the course of the experiment lead to a rejection of \textbf{H3$_0$}. Instead we accepted \textbf{H3$_1$}. Test subjects working with the interactive player executed significantly more optional interactions than their counterparts using a traditional player. 

The results of our study indicate that the concepts we adapted from research on e-learning provide meaningful advantages to the viewing of vision videos. Users of the prototype showed a higher rate of correctly answered questions. This demonstrates that they had a better understanding of the video's content. We attribute this change to the repurposed methods from e-learning contexts.The two groups of participants in our experiment only differed in the video player used.

We also did not find statistically significant differences in the times users spent with the video. Users of both groups of our experiment took similar amounts of time to watch the video. However, users of the prototypical implementation of our concepts gave correct answers more often. We therefore argue that the concepts adapted from e-learning research increase the efficiency with which information is gathered.

The fact that users of the interactive player executed optional interactions more frequently signifies that they were more active than participants in the control group. In accordance with work in the field of educational psychology \cite{chi2014icap} we believe this increase to be the main reason for the positive effects found in our data. The concepts presented in this paper incentivize users to process information actively and lead to increases in the efficiency and clarity of information.

Adopting findings from research on e-learning has lead to meaningful improvements to the use of vision videos. The positive effects of more active viewers could be replicated in our experiment. We were also able to make use of similar methodology to help viewers be more active. The application of concepts from e-learning research to the context of vision videos was successful and should be expanded.




\subsection{Threats to Validity}
We discuss threats to validity according to Wohlin\,et\,al.~\cite{wohlin2012experimentation}.

\paragraph{Conclusion Validity}
The conclusion validity of our results is threatened by the small sample size. An experiment with a larger number of test subjects is required. However, we still found statistically significant results. Another threat is the fact that the presented video directly impacts the results of our study. For example, a more complex vision video complicates answering questions.

\paragraph{Internal Validity}
Questions that were asked over the course of the video could have been too easy or answerable through prior knowledge. This threatens the internal validity. The presence of this threat is indicated by participants answering most questions correctly. Nevertheless, one of the most important aspects of vision videos is their ease of understanding. Therefore, correct answers should be expected, if the shown vision video is of good quality.

\paragraph{Construct Validity}
A threat to the construct validity is posed by participants' lack of practical experience with industrial software projects. The role of a stakeholder has rarely been held by the participants before our experiment. Another threat is that participants who were fully satisfied with the paths they had seen might not want to see more options. In such cases participants see less paths than others who were not happy with the variants they had seen.

\paragraph{External Validity}

The external validity is threatened by the fact that the prototypical video player including the presented concepts did not yet allow users to add annotations directly to the video. However, we would have counted the addition of such annotations as further optional interactions for the purposes of our experiment. Our results already showed a statistically significant increase in optional interactions for the group of participants who used the prototype.



\section{Conclusion}
\label{sec:conclusion}
This paper presents research looking to \textit{engage stakeholders with vision videos through active information processing}. In order to reach this goal, we make use of findings in research on e-learning to design multiple interactions with vision videos. The adapted concepts include questions being asked over the course of a video, as well as alternative paths that viewers can choose from. To evaluate the use of these concepts in the context of vision videos we conducted a study including twelve participants split into two groups. One group used a traditional video player, while the second group was given access to a prototypical implementation of the presented concepts. All participants watched the same video. Our results indicate that the prototype provides clear advantages over a traditional video player. Users gave correct answers to questions more frequently and processed information more actively. We observe advantages of the adapted concepts in the new context that align with the findings in e-learning research. 

Interactivity for the context of vision videos is a research field with various opportunities beyond the concepts presented in this paper. For example, while this paper revolved around the use of vision videos during the negotiation phase, stakeholder feedback is also crucial to the elicitation phase of the requirements engineering process. Here, annotations added by viewers could prove to be a valuable resource of feedback.
Furthermore, we look to study further literature on e-learning and educational psychology to find additional concepts that could support the use of vision videos.
Lastly, a second study consisting of a larger number of participants with more experience of being stakeholders in software projects could lead to more insights into the benefits of interactive vision videos.

\section*{Acknowledgment}
This work was supported by the Deutsche Forschungsgemeinschaft (DFG) under Grant No.: 289386339, project ViViUse.

\bibliographystyle{IEEEtran}
\bibliography{IEEEabrv,references}

\end{document}